\documentclass[english,aps,prb,floatfix,twocolumn,showpacs]{revtex4}
\usepackage[T1]{fontenc}
\usepackage[latin1]{inputenc}
\usepackage{amsmath}
\usepackage{babel}
\usepackage{graphicx}
\usepackage{amssymb}
\usepackage{multirow}

\begin{document}
\title{Evidence of Kinetic Energy Driven Antiferromagnetism in Double Perovskites : A First-principles Study}
\author{Prabuddha Sanyal, Hena Das and T. Saha-Dasgupta}
\affiliation{S.N. Bose National Centre for Basic Sciences,
Kolkata 700098, India}
\date{\today}

\begin{abstract}
Using first principles density functional calculations, together with exact diagonalization of Fe-Mo 
Hamiltonian constructed in a first principles Wannier function basis, we studied the electronic structure 
of La doped double perovskite compound Sr$_2$FeMoO$_6$. Our calculation show stabilization of kinetic energy 
driven antiferromagnetic phase for La rich compounds, in agreement with the results obtained on the basis 
of previous model calculations. 
\end{abstract}
\maketitle
\noindent
\section{Introduction}

Double perovskites with a general formula A$_{2}$BB$^{\prime}$O$_{6}$ where B and B$^{\prime}$ are transition metal ions and A is
a rare-earth or alkaline-earth ion, are materials that have attracted enormous amount of attention in recent
time due to the diversity of their applications as for example in the field of spintronics (Sr$_2$FeMoO$_6$\cite{SFMO}), multiferroicity (Ba$_{2}$NiMnO$_{6}$\cite{BNMO}), magnetodielectric materials (La$_{2}$NiMnO$_{6}$\cite{LNMO}$^,$\cite{LNMO1}), magneto-optic devices (Sr$_{2}$CrOsO$_{6}$, Sr$_{2}$CrReO$_6$\cite{SCOO}). 
 The choice of B and B$^{\prime}$ ions,
provide the tunability of B-O-B$^{\prime}$ interaction, giving rise to a variety of magnetic properties like ferromagnetism,
antiferromagnetism, ferrimagnetism and electronic properties like metallic, half-metallic and insulating~\cite{DDDinesh,DDreview}. The
presence of two transition metal ions instead of one as in perovskite material is expected to give rise to far
more tunability and richness of properties compared to simple perovskites.

Perhaps the most studied member of this series that arose much interest is Sr$_{2}$FeMoO$_{6}$ (SFMO). This material was
reported \cite{SFMO}$^,$\cite{Mag-Res1}$^-$\cite{Mag-Res5}to exhibit a large magnetoresistance(MR) effect with a fairly high ferromagnetic transition temperature of about
410 K, opening up the possibility of designing spintronics materials operating at room temperature. However, unlike colossal magnetoresistive compounds as 
manganites, this MR does not arise from electron-phonon interactions. 
Rather, it is extrinsic, of tunnelling magnetoresistive (TMR) origin. 
 Since the 
report of the large MR effect and high magnetic transition temperature, a number of experimental studies like NMR\cite{kaputza}, XES\cite{kuepper}, Hall
measurements\cite{tomika}, magnetic measurements\cite{topwal} have been carried out to characterize various properties of this material. There have been
also a number of theoretical studies involving both first-principles calculations \cite{TSD}$^-$\cite{Solovyev} as well as model 
calculations \cite{Millis}$^-$\cite{Guinea1}. The unusually high
ferromagnetic transition temperature in Sr$_{2}$FeMoO$_{6}$ and related material like Sr$_{2}$FeReO$_{6}$ was
rationalized \cite{TSD}$^,$\cite{GK} in terms of a kinetic energy driven mechanism which produces a negative spin polarization
at otherwise nonmagnetic site like Mo or Re. Following this idea, a double-exchange like two sublattice model
was introduced and studied by different groups\cite{Millis}$^,$\cite{Avignon}$^,$\cite{Navarro}$^,$\cite{Guinea2}$^,$\cite{Guinea1}.
While most of the studies \cite{Millis}$^,$\cite{Navarro}$^,$\cite{Guinea2} were
restricted only to ferromagnetic phase, some of the studies \cite{Guinea1}$^,$\cite{Avignon} were extended to other competing magnetic
phases too. Very recently \cite{Prabs}, the problem has been studied in detail in terms of a full numerical solution
of spin-fermion model and as well as in terms of reduced, classical magnetic model. These studies predict that
when the competing magnetic phases are taken into account, the electron doped model systems beyond a certain doping prefers to have
antiferromagnetic(AFM) arrangement of Fe spins compared to ferromagnetic(FM) arrangement of the undoped system. {\it The predicted antiferromagnetic phase 
in electron-doped system is kinetic-energy driven rather
than super-exchange driven, as is the case for example in Sr$_{2}$FeWO$_{6}$\cite{SFWO}, which is an insulating antiferromagnet with N\'eel 
temperature of $\approx$ 20 K.} The superexchange driven 
antiferromagnetic phase is necessarily insulating while the kinetic energy driven AFM phase may not be so. The prediction of 
such an antiferromagnetic phase of different origin is therefore of significance. While the kinetic energy driven antiferromagnetic 
phases have been suggested in hole-doped rare-earth manganites (eg.the CE phase at half-doping~\cite{Khomskii}), to the
best of our knowledge, till date no reports of such analogous phases in double perovskites exist, thereby, opening up the possibility 
of experimental exploration in this front. However the afore-mentioned
model calculations were carried out in two dimension and with single band, which was justified by the assumption that the 
dominant nearest-neighbor B-B$^{\prime}$ interactions are operative between orbitals of same symmetry and within a given
plane. These restrictions are not strictly true. Furthermore, the magnetic ordering in real material is three-dimensional.
A full three-dimensional, all orbital calculation without these approximations, 
is therefore necessary to put the possible existence of the AFM phase in firm footing.

Considering the above mentioned points, it is therefore, of interest to study the problem of electron doping using 
first-principles, density functional theory (DFT) based calculations. The first principles
calculations which take into account all the structural and chemical aspects correctly is expected to 
provide more realistic scenario and verification of predictions made by model calculations. The Sr ions 
in SFMO can be substituted for trivalent cations, like La, leading to 
Sr$_{2-x}$La$_{x}$FeMoO$_{6}$. This would cause electron doping in the system, with $1+x$ electron per formula unit
in the conduction band, compared to 1 electron per formula unit in the undoped SFMO situation.  To our knowledge, there exists very few first-principles study of the La-doped SFMO system.
Few studies \cite{JAP}$^,$\cite{ES} that exist explored only the ferromagnetic phase, did not consider the other competing 
magnetic phases and were restricted mostly to Sr-rich part of the phase diagram. Motivated by the findings of the
model calculations \cite{Prabs}, we considered it worthwhile to span the whole concentration range from 
$x=0.0$ {\it ie} Sr$_{2}$FeMoO$_{6}$ to $x=2.0$ {\it ie} La$_{2}$FeMoO$_{6}$ and study the relative stability of the various magnetic
phases as one increases the carrier concentration through the increased doping of La.

We have carried out our study both in terms of full ab-initio calculations as well as 
in terms of solutions of multi-orbital, low-energy Hamiltonians defined in a first-principles derived Wannier function basis.
The structural optimization and total energy calculations of various magnetic phases have been carried
out using the plane wave pseudopotential method as implemented in the Vienna Ab-initio Simulation Package  (VASP) \cite{VASP},
while the doping effect in first-principles calculations has been simulated through
supercell technique. The construction of low-energy Hamiltonian in first-principles derived Wannier function basis 
has been achieved through muffin-tin orbital (MTO)
based Nth order MTO (NMTO)-downfolding technique \cite{NMTO}. The constructed multi-orbital, spin-fermion Hamiltonian defined in the first-principles 
derived Wannier function basis has been
solved by means of real space based exact diagonalization technique.

The rest of the paper is organized in the following manner. Section II contains the details of the employed methods
and calculations. Section III is devoted to results which consist of three subsections: (A) Total energy 
calculations, electronic structure and relative stability of various magnetic phases in doped compounds (B) 
Determination of low-energy, few orbital Hamiltonian by NMTO-downfolding (C) Calculations of 
magnetic phase diagram and magnetic transition temperatures in terms of low-energy Hamiltonian. 
The paper concludes with section IV containing discussion and summary.

\section{Methods and Computational Details}

The first-principles DFT calculations were carried out using the plane wave pseudopotential method implemented within VASP.
We considered exchange-correlation functionals within generalized gradient approximation (GGA) \cite{GGA} and GGA+U \cite{LSDA+U}. 
We used projector augmented wave (PAW) potentials \cite{PAW} and the wavefunctions were expanded in the plane wave basis
with a kinetic energy cut-off of 450 eV. Reciprocal space integration was carried out with a k-space mesh of 6 $\times$ 6 $\times$6.
Two sets of supercell calculations were carried out, one with two formula unit and another with eight formula unit. The
two formula unit supercells with two inequivalent Fe atoms can accommodate the ferromagnetic spin alignment of Fe spins
and the A type antiferromagnetic spin alignments of Fe spins. The eight formula unit supercells with eight inequivalent
Fe atoms in the unit cell, in addition to FM and A type AFM, can accommodate G type antiferromagnetic ordering of Fe 
spins (see Fig.1)

\begin{figure}
\includegraphics[width=9cm]{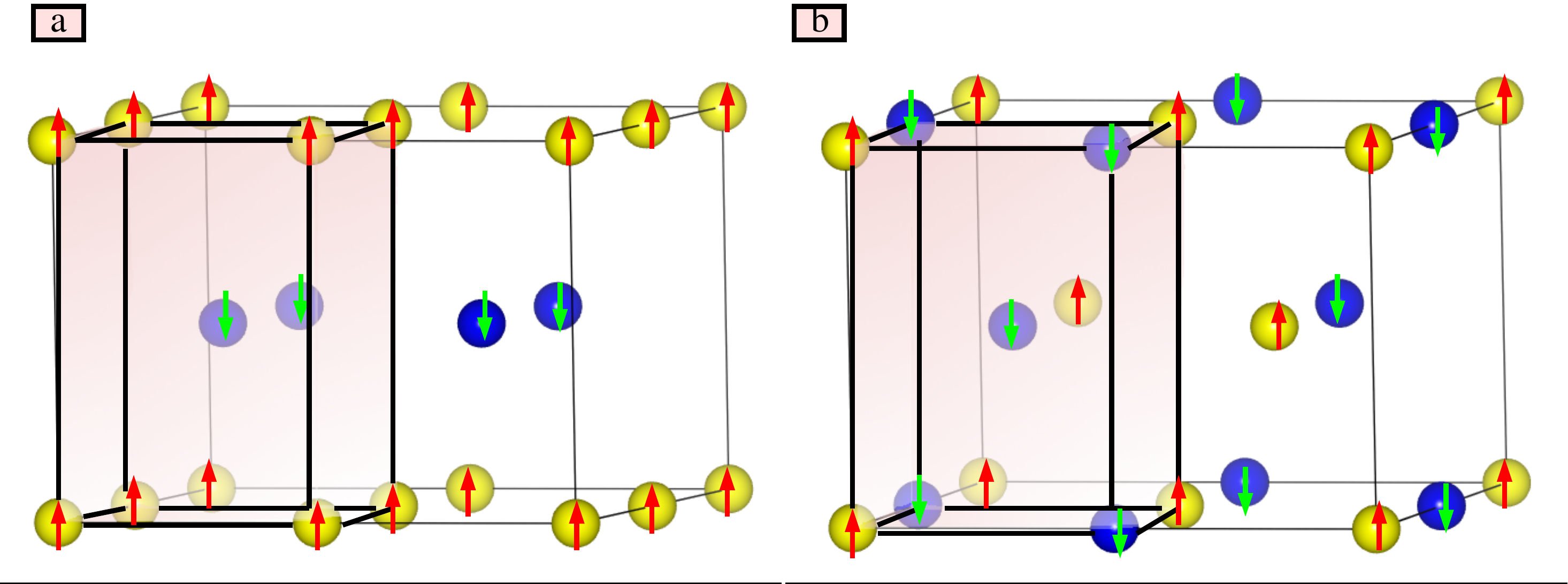}
\caption{\label{fig-1} The Fe sublattice ordering of Sr$_{2-x}$La$_{x}$FeMoO$_{6}$. Shown are the A-type(left panel) and G-type (right panel) antiferromagnetic
arrangement of Fe spins. In case of A-type antiferromagnetism the Fe spins in-plane are ferromagnetically coupled, while
Fe spins between two adjacent plans are antiferromagnetically coupled. For G-type antiferromagnetism, the Fe spins are
antiferromagnetically coupled both out-of-plane and in-plane. The shaded box indicate the unit cell of two formula unit supercell.}
\end{figure}  

For extraction of a few-band, tight-binding Hamiltonian out of full DFT calculation which can be used as input to
multi-orbital, low-energy Hamiltonian based calculations, we have carried out NMTO-downfolding calculations. Starting from a full DFT calculations, 
NMTO-downfolding arrives at a few-orbital Hamiltonian by integrating out degrees 
which are not of interest. It does so by
defining energy-selected, effective orbitals which serve as
Wannier-like orbitals defining the few-orbital Hamiltonian in the {\it
downfolded} representation. NMTO technique which is not yet available in its self-consistent form relies
on the self-consistent potential parameters obtained out of linear muffin-tine orbital (LMTO)\cite{lmto} 
calculations. The results were cross-checked among the plane wave and LMTO calculations in terms of total energy differences, 
density of states and band structures.

The multi-orbital, low-energy Hamiltonian that is assumed to capture the essential physics of SFMO, should consist of the
following ingredients: (i)~a large core spin at the Fe site,
(ii)~strong coupling on the Fe site between the core spin and
the itinerant electron, strongly  preferring  {\it one}
spin polarization of the itinerant electron, and
(iii)~delocalization of the itinerant electron on the Fe-Mo network.

From the above considerations, the representative Hamiltonian is given by:
$$H= \epsilon_{Fe}\sum_{i\in B}f_{i\sigma\alpha}^{\dagger}f_{i\sigma\alpha}+
\epsilon_{Mo}\sum_{i\in B'}m_{i\sigma\alpha}^{\dagger}m_{i\sigma\alpha} $$

$$-t_{FM}\sum_{<ij>\sigma,\alpha}f_{i\sigma,\alpha}^{\dagger}m_{j\sigma,\alpha} 
-t_{MM}\sum_{<ij>\sigma,\alpha}m_{i\sigma,\alpha}^{\dagger}m_{j\sigma,\alpha} $$ 
\begin{eqnarray}
-t_{FF}\sum_{<ij>\sigma,\alpha}f_{i\sigma,\alpha}^{\dagger}f_{j\sigma,\alpha} 
+ J\sum_{i\in A} {\bf S}_{i} \cdot
f_{i\alpha}^{\dagger}\vec{\sigma}_{\alpha\beta}f_{i\beta}
\label{fullhamSFMO}
\end{eqnarray}

The $f$'s refer to the Fe sites and the $m$'s to the Mo sites.
$t_{FM}$, $t_{MM}$, $t_{FF}$ represent the nearest neighbor Fe-Mo, second
nearest neighbor Mo-Mo and Fe-Fe hoppings respectively, the largest hopping being given
by $t_{FM}$.  $\sigma$ is the spin index and $\alpha$ is the orbital index that spans the t$_{2g}$ 
manifold. The difference between the t$_{2g}$ levels of Fe and Mo,
${\tilde \Delta} = \epsilon_{Fe} - \epsilon_{Mo}$, defines the charge transfer energy.
Since among the crystal-field split d levels of Fe and Mo, only the relevant $t_{2g}$ orbitals are retained,
the on-site and hopping matrices are of dimension 3 $\times$ 3. 
The ${\bf S}_i$ are `classical' (large $S$)
 core spins at the B site, coupled
to the itinerant B electrons through a coupling $J \gg t_{FM}$.

Given the fact that  $J \gg t_{FM}$, the Hamiltonian of Eqn~\ref{fullhamSFMO} 
can be cast into form appropriate for  $J\rightarrow\infty$. This gives the following
Hamiltonian, with `spinless' Fe conduction electrons
and Mo electrons having both spin states.
\begin{eqnarray}
H &=& t_{FM}\sum_{<ij>\alpha}
\ ( sin({{\theta_i} \over 2})f_{i\alpha}^{\dagger}m_{j\uparrow\alpha} 
 -
e^{i\phi_i}cos({{\theta_i} \over 2})f^{\dagger}_{i\alpha}m_{j\downarrow\alpha}) \cr
&& +h.c.
+t_{MM}\sum_{<ij>}m_{i\sigma\alpha}^{\dagger}m_{j\sigma\alpha}  \cr
&&
+t_{FF}\sum_{<ij>}cos(\theta_{ij}/2)(f_{i\sigma\alpha}^{\dagger}f_{j\sigma\alpha}) \cr
&&
+\epsilon_{Fe}\sum_{i}f_{i\alpha}^{\dagger}f_{i\alpha}+
\epsilon_{Mo}\sum_{i\sigma\alpha}m_{i\sigma\alpha}^{\dagger}m_{i\sigma\alpha}  
\label{Jinfinityham}
\end{eqnarray}

This is the lowest energy Hamiltonian.
There is no longer any `large' coupling in the Hamiltonian,
and the number of degrees of freedom has been reduced to
three per Fe site and six per Mo, compared to original
problem with six degrees of freedom at both Fe and Mo sites. 
$m_{j\downarrow}$ and $m_{j\uparrow}$
hop to different conduction electron projections at the
neighboring Fe sites so the effective hopping picks up
a $\theta_i, \phi_i$ dependent modulation.
For example, $\theta=0,\phi=0$, corresponds to FM configuration with 
all Fe core spins being up. Since the spin $S$ is large and can be considered 
classical, one can consider different spin configurations (ferro, antiferro and disordered) 
and diagonalize the system in real space, to obtain variational estimates of the ground 
state, and its stability. 

\section{Results}

\subsection{Total Energy, Electronic Structure and Relative Stability of Magnetic Phases}

Sr$_{2}$FeMoO$_{6}$ crystallizes in body centered tetragonal space group with I4/mmm symmetry. The crystal structure
of SFMO is well characterized. The crystal structure of La-doped Sr$_{2}$FeMoO$_{6}$ on the other hand is 
controversial.  Some of the study \cite{P21,p21n} reports that though I4/mmm symmetry is retained for small doping of La, for doping beyond
$x=0.4$ or so, the symmetry changes to P2$_1$/n. The other measurements \cite{JAP} however reports that all compounds
of Sr$_{2-x}$La$_x$FeMoO$_{6}$ for $x=$ 0, 0.25, 0.5 and 1.0 crystallize in I4/mmm symmetry. Unfortunately, the information
of the detail crystal structure data are limited due to the facts that a) the compounds till now have been synthesized
only for La concentrations less than or equal to 1 b) increasing concentration of La leads to increased disorder
which prohibits accurate measurement of the underlying symmetry. While in the following,  
we have primarily reported the results assuming I4/mmm symmetry, we have also carried out calculation
for  P2$_1$/n symmetry for the end member, La$_{2}$FeMoO$_{6}$ (LFMO). The crystal structure corresponding to P2$_1$/n symmetry for La$_{2}$FeMoO$_{6}$
was obtained starting with initial parameters of $x=$ 0.4 as reported in ref.\cite{note} and then performing total energy optimization of 
the initial structure. The P2$_1$/n symmetry structure has been found to be energetically lower in
energy by 90 meV than the corresponding I4/mmm symmetry structure.
However, as described later, the primary conclusion of our results
is found to remain unaffected by this possible change of symmetry.
Table I. shows the
theoretically optimized crystal structures obtained using plane wave basis\cite{note_op}, of SFMO, and that of LFMO assuming I4/mmm symmetry as
well as P2$_1$/n symmetry. 

\begin{table*}
\begin{tabular}{|c|ccc|c|ccc|c|ccc|}
\hline
&\multicolumn{3}{c|}{SFMO}&\multicolumn{8}{c|}{LFMO}\\ 
\hline
& & I4/mmm& & & & I4/mmm& & & & P2$_1$/n & \\
\hline
a& &5.57& &a& & 5.78& &a& & 5.65 & \\
b& &5.57& &b& & 5.78& & b& & 5.63& \\
c& &7.80& & c&& 7.75& &c& & 7.95& \\
 & & & & & & & &$\beta$& & 90.04& \\
 \hline
 & x & y & z && x & y & z & & x & y & z \\
 Sr&0.5&0.0&0.25&La&0.5&0.0&0.25&La&0.010&0.002&0.259\\
O1& 0.248&0.248&0.0&O1&0.245&0.245&0.0&O1&0.504&0.000&0.255\\
O2&0.0&0.0&0.248&O2&0.0&0.0&0.245&O2&0.248&0.257&0.003\\
 & & & & & & & &O3&0.253&0.244&0.497\\
 \hline
 \end{tabular}

\caption{Optimized cell parameters and the atomic positions for Sr$_{2}$FeMoO$_{6}$ and La$_{2}$FeMoO$_{6}$.  Fe and Mo ions are situated at the high symmetry Wykoff positions 2a and 2b, given by (0,0,0) and (0.0, 0.0, 0.5) respectively. For I4/mmm symmetry Sr/La also sites in the high symmetry Wykoff position given by (0.5, 0.0, 0.25) but sits in a general position for P2$_{1}$/n symmetry.}
\end{table*}

The volume for LFMO is found to expand with respect to that of SFMO, in agreement with experimental trend\cite{JAP,P21} of increasing volume with increased La doping. 
 Assuming I4/mmm symmetry, as is seen from Table I, the internal parameters corresponding to oxygen positions, which are the only free parameters within I4/mmm space group, change little upon changing Sr by La. The unit cell volume for various intermediate members of the series obtained by 
interpolation from the optimized lattice parameters of the end members using
Vegar's law,  120.99 \AA $^3$ for SFMO, 123.27 \AA $^3$ for Sr$_{1.5}$La$_{0.5}$FeMoO$_6$ and 125.56 \AA $^3$ for SrLaFeMoO$_6$, agree well with the experimental data available for I4/mmm symmetry in terms of volume expansion, given by 121.4 \AA $^3$, 124.0 \AA $^3$ and 124.88 \AA $^3$ respectively\cite{JAP}. The crystal structure for the doped compounds in the assumed I4/mmm symmetry for the intermediate concentration values are, therefore, obtained by
using Vegard's law for interpolation of cell parameters keeping the atomic positions fixed.

In the next step, we have carried out total energy calculations of Sr$_{2-x}$La$_{x}$FeMoO$_6$ in I4/mmm symmetry for the FM 
alignment of Fe spins and the AFM alignment of Fe spins, which for a two formula unit supercell is of A type 
(see Fig. 1). The energy difference between FM and AFM-A spin configuration per formula unit as a function of La concentration is
plotted in Fig. 2. Calculations have been carried out both within GGA and GGA+U. Focusing on to GGA results first, as is evident from Fig. 2, the stability of the FM phase with respect to AFM configuration is
gradually reduced as the La concentration is increased. As the concentration is increased beyond $x$=1.5 or so, the FM
phase becomes unstable and the AFM phase becomes the ground state, in agreement with prediction of 
model calculations \cite{Prabs}$^,$\cite{Guinea1}. The total and magnetic moments at Fe and Mo sites, as obtained within GGA, 
are listed in Table II. The net magnetic
moment at the FM phase reduces as the La concentration is increased, which is due to the increased moment at the Mo
site. Such behavior has been also observed in experiment \cite{JAP}. Especially, photoemission studies have confirmed that electron injection occurs at the Mo site, increasing the moment
 on that site~\cite{photoemission}. While the moment at the Fe site stays more or less the same
between ferromagnetic and antiferromagnetic phase, the magnetic moment at the Mo site is found to be systematically 
smaller in the AFM phase compared to FM phase.

In order to check the influence of the possible change of crystal symmetry that may happen between SFMO and LFMO, we have also
calculated the total energy difference between FM and AFM-A spin configurations, assuming LFMO in P2$_{1}$/n symmetry with
theoretically optimized structure. The calculated E$_{FM}$ - E$_{AFM-A}$ came out to be 0.094 eV per formula unit, confirming the stabilization of
AFM phase for LFMO. While the possible change of crystal symmetry from I4/mmm to P2$_{1}$/n for La rich samples is expected
to change the precise La concentration at which FM to AFM transition happens, the general trend of AFM phase becoming
progressively more favorable upon increasing La doping therefore would remain hold good.

Fig. 3 shows the GGA density of states corresponding to FM phase of SFMO, LFMO and the doped compounds, SrLaFeMoO$_{6}$
and Sr$_{0.5}$La$_{1.5}$FeMoO$_{6}$ in I4/mmm symmetry. Focusing on the well-studied \cite{TSD} DOS of SFMO, we find that the Fe $d$ states 
are nearly full (empty) in the majority (minority) spin channel while the Mo $d$ states are nearly empty in the majority
spin channel and partially filled in the minority spin channel. This is in conformity with the half metallic character
of the compound and also with the nominal Fe$^{3+}$ and Mo$^{5+}$ valences. Due to the octahedral oxygen surrounding
of Fe and Mo atoms, the Fe $d$ and Mo $d$ states are split up into t$_{2g}$ and e$_g$, the highly delocalized state
crossing the Fermi level in the minority spin channel being of mixed Fe-t$_{2g}$-Mo-t$_{2g}$ character. The empty Mo-t$_{2g}$ states in the majority spin channel is found to be highly localized giving rise to peaked structure positioned at about 1 eV above the Fermi energy. As each of the Sr atoms is replaced by a La atom, one extra electron is introduced in the system which populates the hybridized Fe-t$_{2g}$-Mo-t$_{2g}$ state in the minority spin channel, keeping the overall structure of the density of states
intact. The Fermi level therefore progressively moves up like a rigid band fashion as $x$ is increased and eventually hits the
van Hove singularity of the Mo-t$_{2g}$ states in the majority spin channel. The FM solution becomes unstable at this point. This is schematically
shown in the left panel of Fig. 5. Interesting the DOS corresponding to the mixed Fe-t$_{2g}$-Mo-t$_{2g}$ character in the minority spin
channel also exhibits the singularity at the same energy due to the essentially 2D-like nature of the hoppings between Mo-t$_{2g}$ and
Fe-t$_{2g}$ Wannier functions as will be discussed in the following section.

\begin{figure}
\includegraphics[width=6cm,angle=0]{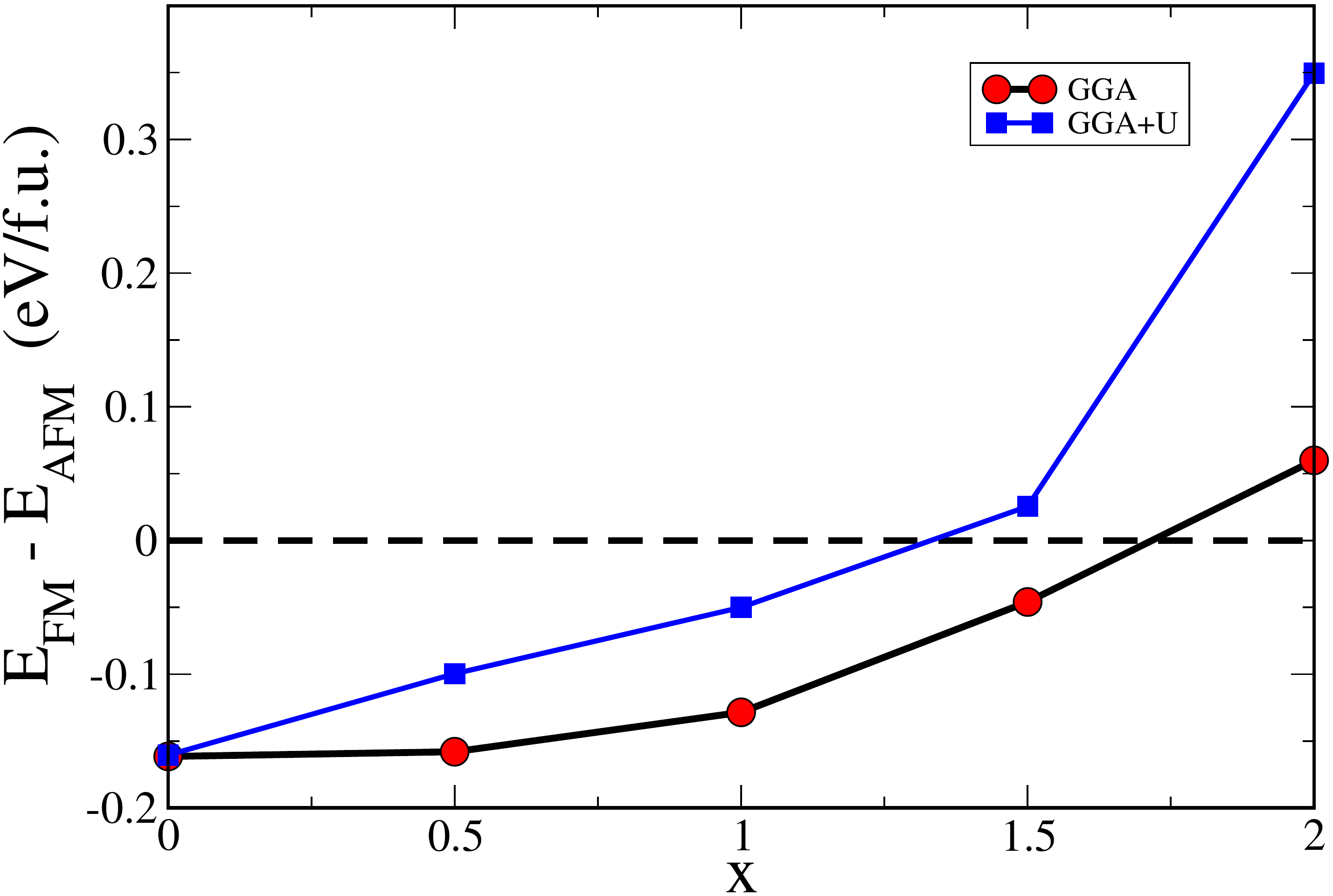}
\caption{\label{fig-2} The energy difference between FM and AFM-A phase plotted as a function of La concentration. 
The FM phase becomes unstable beyond a critical concentration of La both within GGA and GGA+U.}
\end{figure}  

\begin{table}
\begin{tabular}{|c|c|c|c|c|c|c|}
\hline 
&&SFMO&S3LFMO&SLFMO&SL3FMO&LFMO\\
\hline
\multirow{3}{*}{FM}&Fe &3.68&3.59&3.53&3.50&3.52\\
\cline{2-7}
&Mo&-0.23&-0.45&-0.71&-0.80&-0.85\\
\cline{2-7}
&Total&4.0&3.5&3.0&2.5&2.0\\
\hline
\hline
\multirow{3}{*}{AFM}&Fe&3.69&3.60&3.52&3.42&3.50\\
\cline{2-7}
&Mo&-0.05&-0.03&-0.04&-0.18&-0.70\\
\cline{2-7}
&Total&0.0&0.0&0.0&0.0&0.0\\
\hline 
\end{tabular}
\caption{Magnetic moments at Fe and Mo sites, and the total magnetic moment in FM and AFM-A phase of 
Sr$_{2-x}$La$_{x}$FeMoO$_{6}$ in a two formula unit calculation. S3LFMO, SLFMO, SL3FMO refer to Sr$_{1.5}$La$_{0.5}$FeMoO$_6$,
 SrLaFeMoO$_6$ and Sr$_{0.5}$La$_{1.5}$FeMoO$_6$ respectively.}
\end{table}

\begin{figure}
\includegraphics[width=8cm,angle=0]{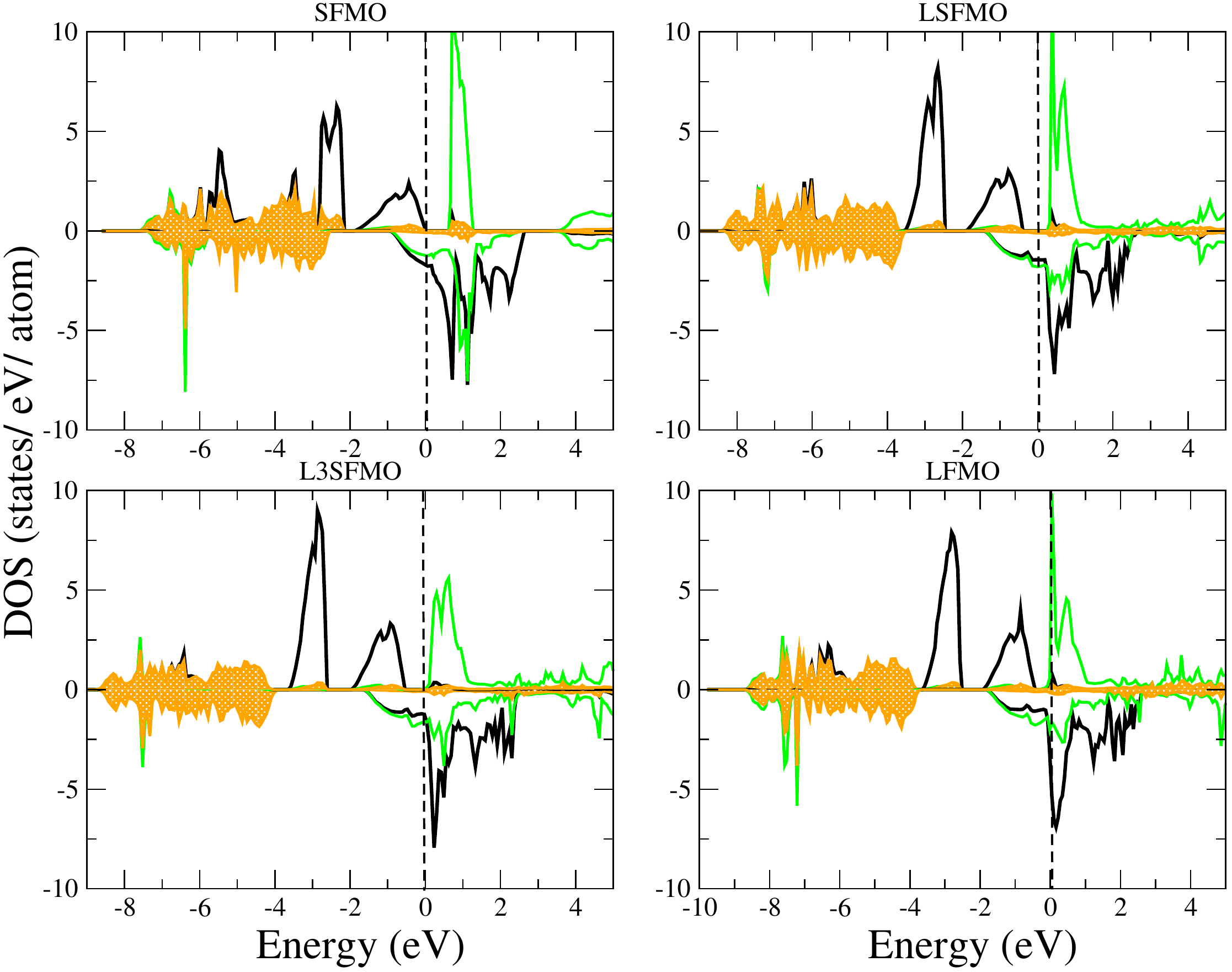}
\caption{\label{fig-3} (Color online) The GGA density of states corresponding to SFMO, SrLaFeMoO$_6$ (SLFMO), 
Sr$_{0.5}$La$_{1.5}$FeMoO$_6$ (L3SFMO) and LFMO in ferromagnetic configuration. The density of
states projected onto Fe, Mo and O are represented by solid black, green (grey) and shaded grey area. The upper and
lower panels correspond to majority and minority spin channels. Zero of the energy is set at the GGA Fermi energy.}
\end{figure}  

Fig. 4 shows the density of states of SFMO, LFMO and the doped compounds, SrLaFeMoO$_{6}$
and Sr$_{0.5}$La$_{1.5}$FeMoO$_{6}$ in the antiferromagnetic A phase, as calculated within GGA. In the two formula unit
supercells, there are two inequivalent Fe atoms, Fe1 and Fe2, whose spins are antiferromagnetically
oriented. The majority channel of Fe1 therefore is identical to the minority channel of Fe2 and vice versa.
The induced moments at two inequivalent Mo sites also become antiferromagnetically aligned, giving rise
to a net AFM arrangement with a zero total moment. Shown in Fig 4, are therefore, the partial DOS corresponding
to one of the sublattice since that of the other sublattice is identical with majority and minority spins reversed.
We find that the Mo-Fe hybridized state crossing the Fermi level, has a three peak van Hove structure. This arises because
of the fact that due to creation of sublattices in the AFM phase, the Mo hopping becomes restricted to a reduced
dimension as the Mo electrons can effectively hop to Fe sites with a specific orientation of Fe spins and not
in another. Interestingly, such a three peak structure formation is also seen in model calculation (see Fig-3 of ref[28]). As found
in the case of FM DOS, the gross features of the density of states remain unchanged with the La doping apart from
the upward shift of the Fermi energy. Reaching LFMO, the Fermi level lands up in the dip of the three peak structured
DOS, justifying the stability of the antiferromagnetic phase, as shown in the schematic diagram of Fig. 5.

The antiferromagnetic state becomes energetically favorable, when the filling is such that it starts populating the 
Mo states in the majority spin channel of the FM DOS, which is highly localized due to the strong preference of the Mo-Fe hopping in one spin channel
and not in another. The antiferromagnetic configuration of Fe spins, on the other hand, allows both Mo
down spin as well as up spin electron to hop, albeit in different sublattices, thereby stabilizing the AFM phase through kinetic energy gain.

\begin{figure}
\includegraphics[width=8cm,angle=0]{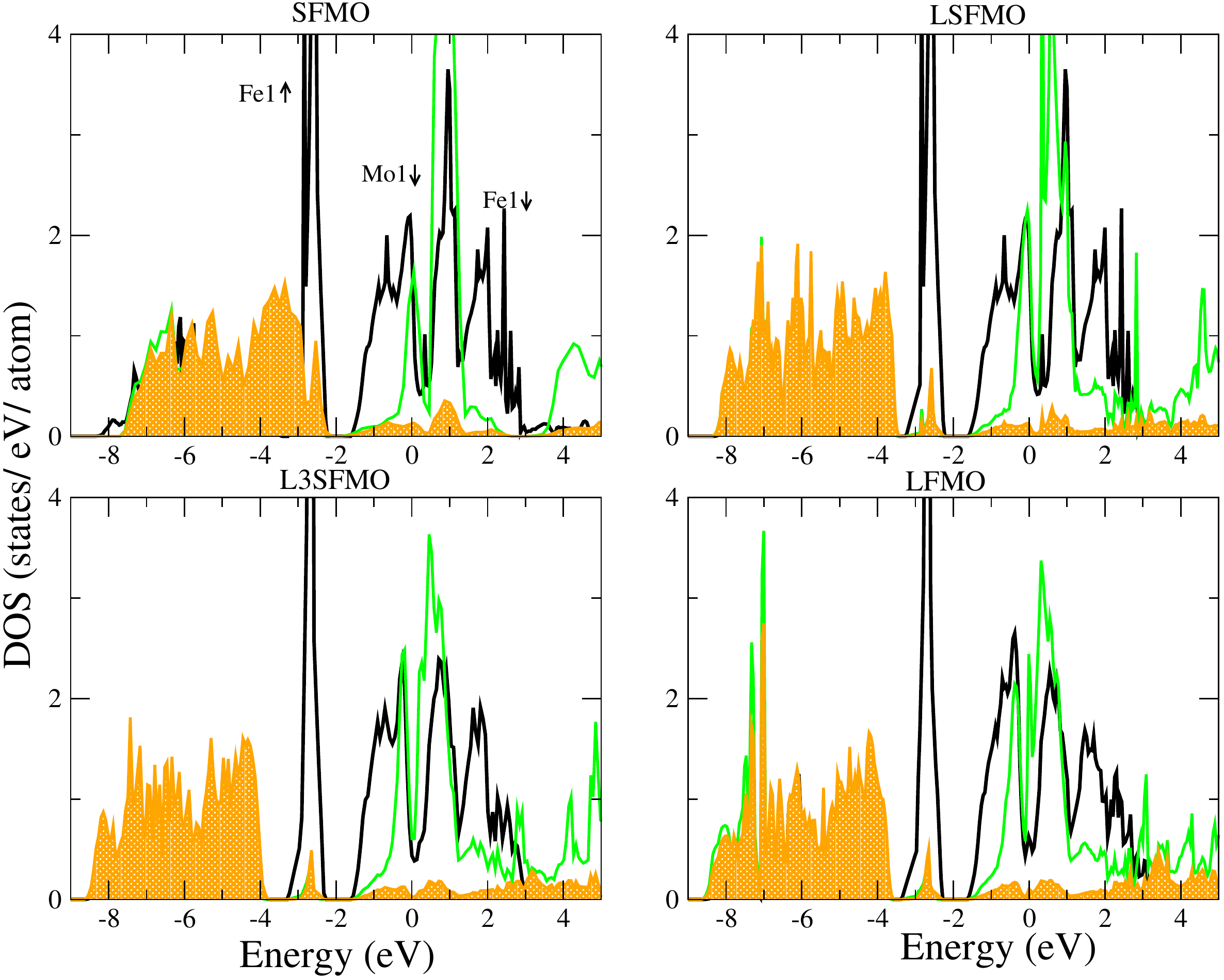}
\caption{\label{fig-4} (Color online) The density of states corresponding to SFMO, SrLaFeMoO$_6$ (SLFMO), 
Sr$_{0.5}$La$_{1.5}$FeMoO$_6$ (L3SFMO) and LFMO in the A-type antiferromagnetic configuration. The density of
states projected onto Fe, Mo and O are represented by solid black, green (grey) and shaded grey area. Zero of the energy is 
set at the GGA Fermi energy.}
\end{figure}

In order to check the influence of the missing correlation effect in GGA, we have also carried out GGA+U calculations
with a typical U value\cite{uvalue} of 4 eV and J value of 1 eV, applied at the Fe site. The calculated energy difference between FM and AFM-A configuration
as a function of La doping is shown in Fig 2, along with GGA results. The application of U is found to increase the relative
stability of AFM phase due to the  increased superexchange contribution to antiferromagnetism in addition to kinetic energy
driven antiferromagnetism.

In Fig~\ref{fig-LDAU}, we show the GGA+U DOS for LFMO, plotted for both FM and AFM-A phases. It is observed
that the gross features of the DOS close to Fermi energy, remain similar to GGA: in particular, the Fermi energy in the FM phase
 remains pinned to the unoccupied Mo $t_{2g}$ level in the FM phase. However, the hybridization
 between the Fe and Mo decreases. Nevertheless, the antiferromagnetic state is still found to have a finite density of
states at Fermi energy, signifying the dominance of kinetic energy driven contribution over that of superexchange.

\begin{figure}
\includegraphics[width=8cm,angle=0]{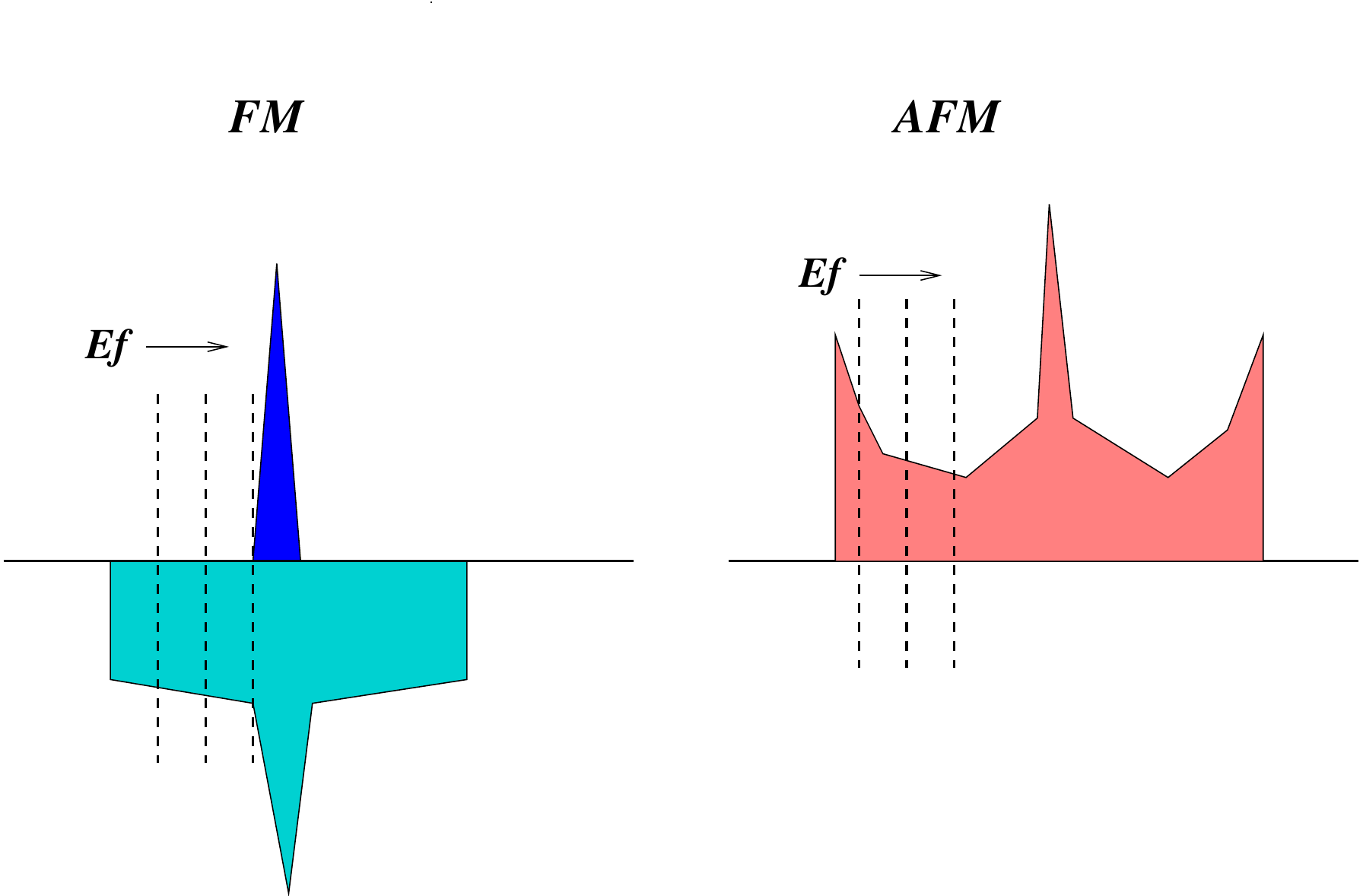}
\caption{\label{fig-5} (Color online) Schematic diagram showing the mechanism stabilizing the AFM phase over FM phase. As the
La doping is increased, the Fermi level (E$_f$) shifts towards right.}
\end{figure}  

\begin{figure}
\includegraphics[width=8cm,angle=0]{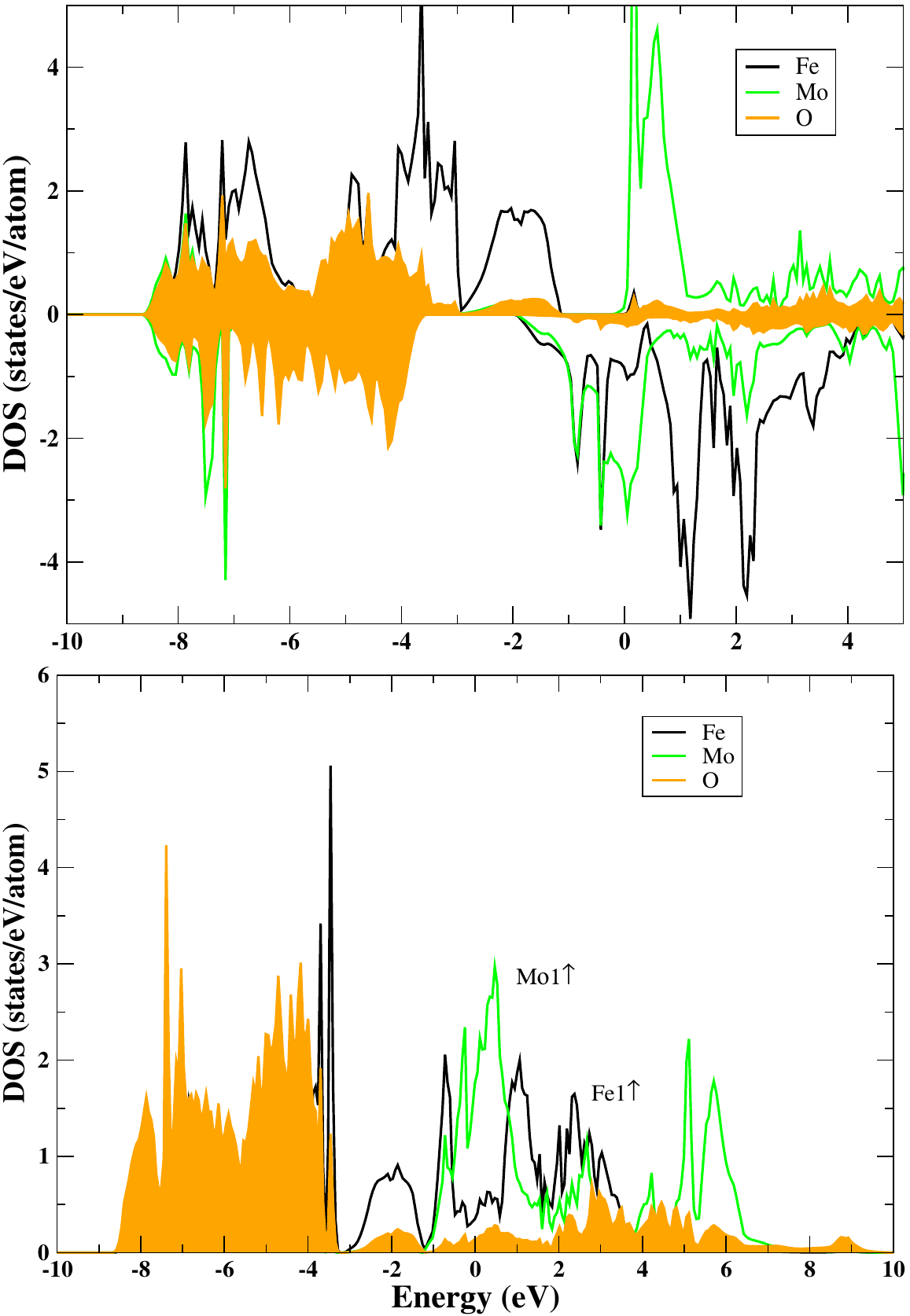}
\caption{\label{fig-LDAU} (Color online){DOS for LFMO in FM and AFM-A phase, using GGA+U}}  
\end{figure}

As already mentioned, considering the two formula unit supercell, the possible AFM arrangement that can be achieved is
of A type. In order to achieve the G type AFM ordering involving both in-plane and out-of-plane AFM ordering, one
needs to increase the size of the supercell to at least eight formula unit. Eight formula unit supercells also allow to
probe the concentration range intermediate to $x$=1.5 and $x$=2.0, the region where the crossover from FM to AFM
happens. Since the qualitative conclusions remain unchanged between GGA and GGA+U, the eight formula unit calculations
were carried out only for GGA.
The energy differences per formula unit obtained for different concentrations of La between FM and AFM-A, and 
between FM and AFM-G phases
are listed in Table III. As found in the calculations with two formula unit, the stability of the FM phase is found
to gradually decrease as the La concentration increases. Among the two antiferromagnetic phases, the G-type
AFM is found to be energetically very close to A type AFM phase, with G type AFM being the ground state at the
end limit of doping {\it i.e.} for LFMO.

\begin{table}
\begin{tabular}{|c|c|c|}
\hline
 & $\Delta$E(FM - AFM-A) &$\Delta$E(FM - AFM-G) \\
\hline
SFMO &-0.145  &-0.147  \\
SLFMO & -0.076 & -0.073 \\
Sr$_{0.5}$La$_{1.5}$FeMoO$_6$ & -0.017 & -0.008 \\
Sr$_{0.375}$La$_{1.625}$FeMoO$_6$&0.014&0.006\\
Sr$_{0.25}$La$_{1.75}$FeMoO$_6$&0.037&0.032\\
Sr$_{0.125}$La$_{1.875}$FeMoO$_6$&0.057&0.052\\
LFMO & 0.066  & 0.069 \\
\hline

\end{tabular}
\caption{Total energy differences per unit formula in eV between FM and AFM-A, and between FM and AFM-G
for various doping of La, as obtained within eight formula unit supercell calculations.}
\end{table}

\subsection{Determination of low-energy, few orbital Hamiltonian by NMTO-downfolding}

\begin{figure}
\includegraphics[width=8cm,angle=0]{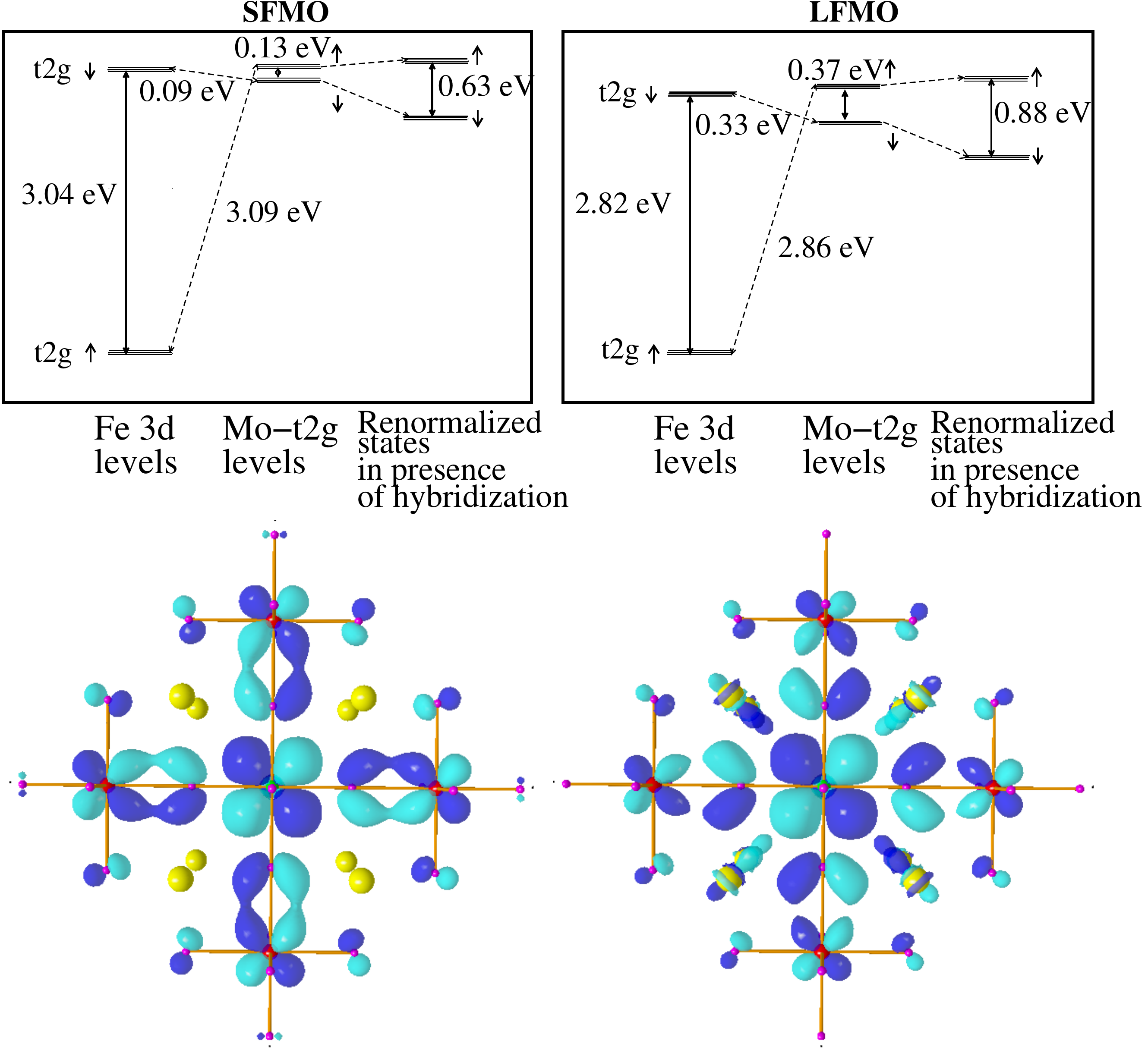}
\caption{\label{fig-6} (Color online) Top panels: Positioning of various energy levels as obtained by NMTO-{\it downfolding}
calculation before and after switching on the hybridization between the magnetic and nonmagnetic ions. Bottom panels: Effective Mo-t$_{2g}$ 
Wannier orbitals corresponding to massively downfolded NMTO Hamiltonian
in the down spin channel. Shown are the orbital shapes (constant-amplitude surfaces) with lobes of opposite
signs colored as  blue (dark grey) and cyan (light grey). The central part of the Wannier orbitals are shaped according to Mo-t$_{2g}$ symmetry, while
the tails are shaped according to Fe-t$_{2g}$ and O-p symmetries. Significant amount of weights are seen in O and Fe site which reflects the strong hybridization
between Fe, Mn and O. For LFMO, finite weights are seen also at La sites, occupying
the hollows formed between Mo-O and Fe-O bonds, which is of La 3z$^{2}$ character.}
\end{figure}

In order to probe the variation of La concentration in a continuous manner, it is perhaps more convenient to adopt a 
low-energy Hamiltonian approach. This would also allow one to calculate the physical properties like magnetic transition
temperatures, transport, spin wave spectra in a much more manageable way. For this purpose, however, it is essential
to construct a realistic, low-energy Hamiltonian. We have used for this purpose, the NMTO-downfolding technique. For 
the present problem, we have derived a Fe-t$_{2g}$ - Mo-t$_{2g}$ only Hamiltonian by
integrating out all the degrees of freedom other than Fe-t$_{2g}$ and Mo-t$_{2g}$. Calculations were carried out
both in the spin-polarized and non spin-polarized form. First of all, Fig.~\ref{fig-6} illustrates the driving mechanism of
magnetism in this class of compounds\cite{tl2mn2o7}. The top panels show the on-site energies of the real-space Hamiltonian defined
in downfolded effective Fe-Mo basis for SFMO and LFMO in a spin polarized calculation. As is seen, the
t$_{2g}$ levels of Mo appear in between the exchange split Fe d states. Upon switching on the hybridization between
Fe-d and Mo-t$_{2g}$, states of same symmetry and spin interact. As a result, Mo-t$_{2g}$ up spin states are pushed
up in energy and Mo-t$_{2g}$ down spin states are pushed down in energy, introducing a renormalized, negative spin
splitting at the Mo site. The normalized spin splitting at Mo site is estimated by massive downfolding procedure
by keeping only Mo-t$_{2g}$ states active in the basis, as shown in the right half on the top panels in Fig.~\ref{fig-6} . We
note that this to be true for both SFMO and LFMO. This in turn, once again, reconfirms the hybridization
driven mechanism to be operative both in SFMO and LFMO, the only difference
 being in the carrier concentration. This is in contrast to Sr2FeMoO6
where W t2g-O hybridized levels are pushed above the exchange split Fe d levels. The
increase in the number of conduction electrons for LFMO compared to SFMO, is reflected in the spin splitting
at Mo site before switching of the hybridization, to be about three times larger in LFMO (0.37 eV) compared to
that of SFMO (0.13 eV). The bottom panels of Fig.~\ref{fig-6} show the plots of Wannier functions of the massively downfolded
Mo-t$_{2g}$ in the down spin channel which demonstrates the hybridization between Mo-t$_{2g}$ and Fe-t$_{2g}$ states.

                     Table IV. shows the hopping interactions between Fe and Mo, obtained in the basis of Fe and Mo t$_{2g}$ Wannier functions constructed by NMTO-downfolding technique. The numbers inside the bracket
are that of LFMO while those outside are that of SFMO.  The examination of the hopping table reveals that the nearest neighbor Fe-Mo hopping to be strongest, as expected. The second nearest neighbor Mo-Mo hopping is half as strong as the nearest neighbor Fe-Mo hopping, while the second nearest neighbor Fe-Fe hopping is about one fifth of the nearest neighbor Fe-Mo hopping. The out-of-plane hoppings which are of dd$\delta$ kind are order of magnitude smaller than the in-plane dd$\pi$ kind of hopping, while inter-orbital hoppings are found to be negligibly small (less than 0.01 eV). 
This makes the hopping essentially two dimensional, as commented earlier.
As is also evident, by replacing Sr by La, the essential 
material specific parameters of the low-energy Hamiltonian, as given in Eqn. 1 changes very little. This is shown pictorially in Fig~\ref{fig-6}, where it is found that
the relative energy positions of the $t_{2g\downarrow}$ levels of Fe and Mo change very
little in going from SFMO to LFMO. In the solution
of low-energy Hamiltonians, to be described in the next section, calculations are therefore carried out
assuming the hopping parameters corresponding to SFMO and varying the carrier concentration in a rigid-band
fashion. The charge transfer energy between Fe-t$_{2g}$ and Mo-t$_{2g}$ has been found to differ by about 5$\%$ which
has been taken into account in these calculations.

\begin{table}[h]
\begin{tabular}{|c|c|c|c|c|}
\hline
&Direction &xy,xy&yz,yz&xz,xz\\
\hline
\multirow{3}{*}{1NN }&[100]&-0.26 (-0.26)&-0.02 (-0.04)&-0.26 (-0.26)\\
&[010]&-0.26 (-0.26)&-0.26 (-0.26)&-0.02 (-0.04)\\
(Fe-Mo)&[001]&-0.02 (-0.04)&-0.26 (-0.25)&-0.26 (-0.25)\\
\hline
\multirow{3}{*}{2 NN }&[110]&-0.05 (-0.06)&0.01 (0.00)&0.01 (0.00)\\
&[101]&0.00 (0.00)&0.01 (0.00)&-0.04 (-0.06)\\
(Fe-Fe)&[011]&0.00 (0.00)&-0.04 (-0.06)&0.01 (0.00)\\
\hline
\multirow{3}{*}{2 NN }&[110]&-0.11 (-0.12)&0.00 (0.01)&0.00 (0.01)\\
&[101]&0.01 (0.01)&0.00 (0.01)&-0.11 (-0.12)\\
(Mo-Mo)&[011]&0.01 (0.01)&-0.11 (-0.12)&0.00 (0.01) \\
\hline
3NN &[111]&0.01 (0.00)&0.00 (0.00)&0.00 (0.00)\\
(Fe-Mo)&&&&\\
\hline
\multirow{3}{*}{4 NN }&[100]&0.01 (0.01)&0.01 (0.00)&0.01 (0.01)\\
&[010]&0.01 (0.01)&0.01 (0.01)&0.01 (0.00)\\
(Fe-Fe)&[001]&0.01 (0.00)&0.01 (0.01)&0.01 (0.01)\\
\hline
\multirow{3}{*}{4 NN }&[100]&0.01 (0.03)&0.01 (0.00)&0.01 (0.03)\\
&[010]&0.01 (0.03)&0.01 (0.03)&0.01 (0.00)\\
(Mo-Mo)&[001]&0.01 (0.00)&0.01 (0.03)&0.01 (0.03)\\
\hline
\multirow{3}{*}{5 NN }&[110]&-0.01 (-0.01)&0.00 (0.01)&0.00 (0.00)\\
&[101]&0.00 (0.01)&0.00 (0.00)&-0.01 (-0.01)\\
(Fe-Mo)&[011]&0.00 (0.00)&-0.01 (-0.01)&0.00 (0.01)\\
\hline
\end{tabular}
\caption{Hopping matrix elements in eV between Fe-t$_{2g}$ and Mo-t$_{2g}$. Only the hopping matrix elements of magnitude larger than 0.01 eV are listed.
The onsite matrix elements are given by 0.005 (0.008) eV, 0.0 (0.0) eV, 0.0 (0.0) eV for Fe-$xy$, Fe-$yz$ and Fe-$xz$ respectively, and 
1.018 (1.057) eV, 1.007 (1.053) eV, 1.007 (1.053) eV for Mo-$xy$, Mo-$yz$ and Mo-$xz$ respectively. All numbers inside the bracket are 
for LFMO and those outside are for SFMO. The energies for a given compound is measured with respect to the lowest energy state. The small
differences between numbers involving $xy$ and that of $yz$ and $xz$ reflect the tetragonality present in the systems.}
\end{table}

\subsection{Calculations of Magnetic Phase Diagram and Magnetic
Transition Temperatures in terms of Low-energy Hamiltonian}

The exact diagonalization of the low-energy Hamiltonian, as given in Eqn. 2 has been carried out for finite size lattice of dimensions
4 $\times$ 4 $\times$ 4, 6 $\times$ 6 $\times$ 6 and 8 $\times$ 8 $\times$ 8. The hopping parameters and the onsite
energies were taken out of DFT calculations, as listed in Table IV. For convenience of calculation, we have neglected 
the small tetragonality reflected in the parameters
listed in Table IV. The dominant hopping interaction which is between
nearest-neighbor Fe and Mo is found to be of the order of 0.3 eV, while the spin exchange splitting at Fe site
as shown in Fig.~\ref{fig-6}, is of order of the 3 eV, an order of magnitude larger than the dominant hopping interaction.
This justifies the assumption of  $J\rightarrow\infty$ limit as adopted in Eqn. 2. 
This makes the rank of the Hamiltonian to be diagonalized as 9/2 $\times$ N$^{3}$ for
a N $\times$ N $\times$ N lattice\cite{footnote}. 

\begin{figure}
\includegraphics[width=7cm]{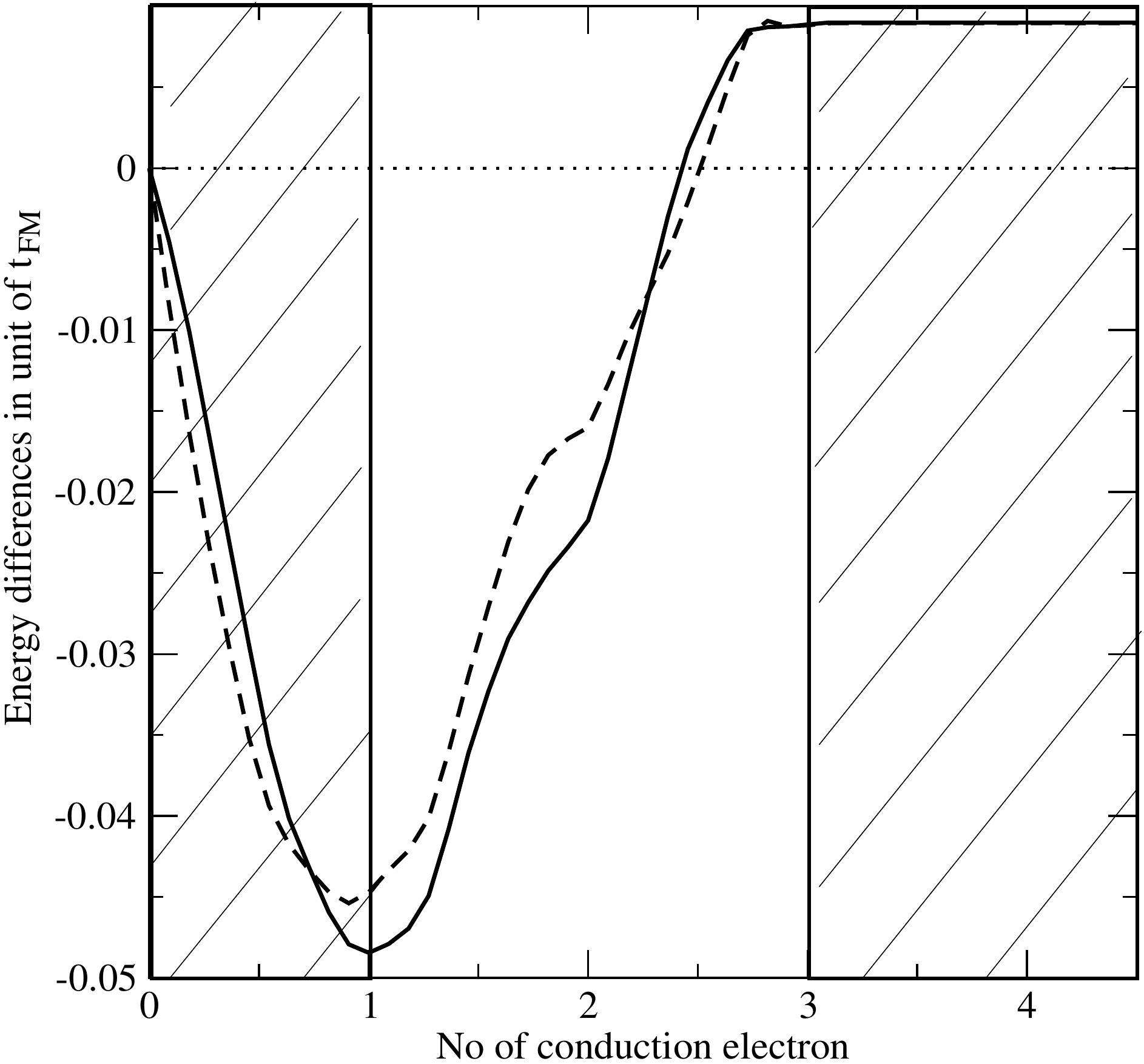}
\caption{\label{fig-7} The energy differences between the FM and G-type AFM phase (solid line) and the FM and A-type AFM 
phase (dashed line) plotted as a function of the number of conduction electrons, as obtained by exact diagonalization of
the low-energy Hamiltonian for a 8 $\times$ 8 $\times$ 8 lattice. Only the region outside the hashed regions, from carrier
concentration 1 to 3 is of relevance for Sr$_{2-x}$La$_{x}$FeMoO$_{6}$.}
\end{figure}

The energy difference between ferromagnetic configuration and G-type and A-type antiferromagnetic configuration of Fe spins
as a function of carrier concentration is plotted in Fig.~\ref{fig-7} . The negative values of the energy differences indicate the
stability of the ferromagnetic phase while the positive values indicate the stability of the antiferromagnetic phase.
The cross-over happens for a value of conduction electrons equal to about $\sim$ 2.6, corresponding to $x=$ 1.6, which agrees well with the results of
8 formula unit supercell calculations, given the assumption of infinite Hund's coupling at Fe site and the finite size effect.
This agreement is nontrivial, since the effective
Hamiltoanian has only 12 spin-orbitals, and hence 12 bands, as compared to the 500 band
calculation with 8-formula unit supercells. This in turn, validates the construction of low-energy model Hamiltonian as given in Eqn. (2), in terms of correct 
identification of the essential contributing terms. This gives us confidence in the constructed
low energy model Hamiltonian, which can henceforth be used to calculate many other
properties like conductivity, susceptibility, magnetoresistance, including at finite temperature, which are not easily accessible within DFT.

\begin{figure}
\includegraphics[width=7cm,angle=0]{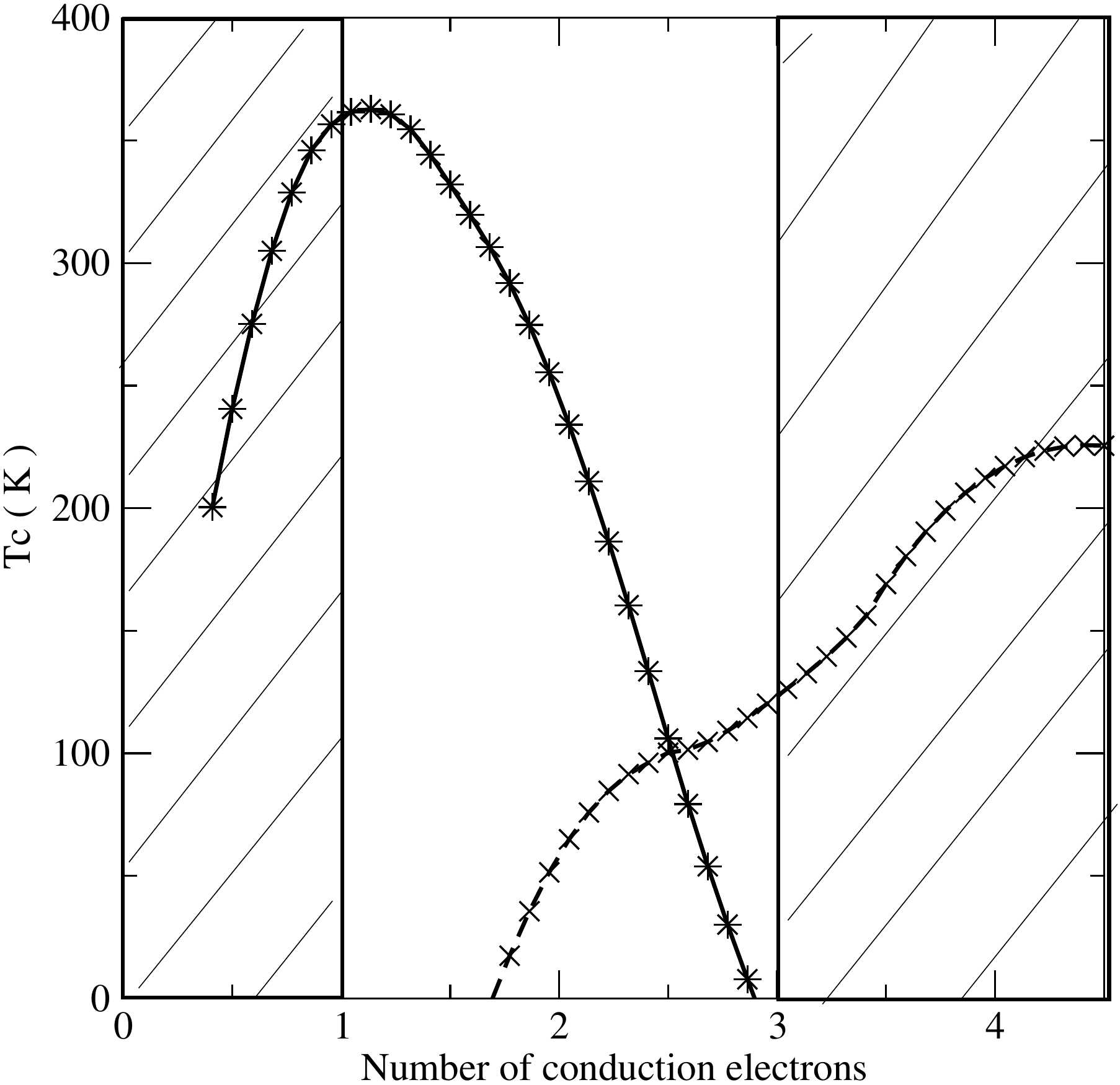}
\caption{\label{fig-8} The ferromagnetic T$_c$ (solid line) and the antiferromagnetic transition temperature T$_N$ (dashed line)
plotted as a function of the number of conduction electrons, as obtained by exact diagonalization of
the low-energy Hamiltonian for a 8 $\times$ 8 $\times$ 8 lattice. As in Fig~\ref{fig-7} , the region outside the hashed regions, from carrier
concentration 1 to 3 is of relevance for Sr$_{2-x}$La$_{x}$FeMoO$_{6}$.}
\end{figure}

As an example, we have used the solutions of the low-energy Hamiltonian to calculate the magnetic transition temperatures by calculating
the difference between the paramagnetic phase and the relevant magnetic phases. The paramagnetic phase
was simulated as disordered local moment calculations, where the calculations were carried out for several 
( $\sim$ 50) disordered configurations of Fe spin and were averaged to get the energy corresponding to paramagnetic phase.
We note that, such a calculation would have been rendered extremely difficult 
within ab-initio
owing to the computational time involved using large supercells, and also averaging them
over myriad configurations.
Fig.~\ref{fig-8}  shows the transition temperatures plotted as a function of carrier concentration. The ferromagnetic transition
temperature at carrier concentration of 1, which corresponds to SFMO compound, is found to be 360 K in comparison
to measured value of 410 K \cite{Delin}. The ferromagnetic T$_c$ is found to decrease upon increasing La concentration, and finally
becomes zero. Upon suppression of ferromagnetic T$_c$, the transition temperature of the antiferromagnetic phase, T$_N$ starts
growing, hitting a maximum value for the end member, LFMO.

\section{Summary and Outlook} 

Using the combination of first-principles DFT calculations and exact diagonalization calculations of low-energy Hamiltonians, 
we showed that the La doped Sr$_{2}$FeMoO$_{6}$ compounds become progressively more unstable towards ferromagnetism as the La
concentration is increased. For the La rich members of Sr$_{2-x}$La$_x$FeMoO$_6$ series with $x > 1.6$, 
the ground state becomes antiferromagnetic. This antiferromagnetic
phase is found to be governed by the kinetic energy driven mechanism as operative in SFMO and achieved by 
change in carrier concentration. In contrast to the super-exchange driven antiferromagnetic phase found in
case of double perovskite like Sr$_{2}$FeWO$_6$ \cite{SFWO}, this antiferromagnetic phase presumably is metallic.
Our DFT calculations found antiferromagnetic solutions with finite density of states at Fermi energy. The preliminary
calculations of the matrix elements of the current operator also turned out to be non-zero. This will be
taken up with more rigor in near future.

It is interesting to compare our results with Dynamical Mean Field Theory (DMFT) calculations done by Chattopadhyay and Millis~\cite{Millis}, 
using a one-band model Hamiltonian. This was, however, a single site calculation, and hence there was no possibility of
capturing an antiferromagnetic phase. Their $T_{c}$ vs $N$ plot for the ferromagnetic phase,
however, was very similar to ours, as shown in Fig. 9, in the sense that the $T_{c}$ first increased, and then decreased
with increasing filling, finally, becoming $0$ at a filling close to $3$. The additional and the most important finding
of our study is that  our calculations also demonstrate the cause of the vanishing $T_{c}$: namely the emergence of the AFM phase.

There are however, several important issues which needs to be considered. Formation of pure La$_{2}$FeMoO$_{6}$ to
best of our knowledge is not been reported in literature, which presumably is due to relative scarcity of Mo$^{3+}$
ions in octahedral environment. La rich SFMO samples, therefore seem more promising candidates for exploration
of the antiferromagnetic phases. 

Another important issue to bother about is the issue of antisite disorder, which has not been considered in our study.
The necessary conditions of formation of double perovskites with ordered, rock salt arrangement of B and B$^{\prime}$ 
transition metal ions are that the size difference between  B and B$^{\prime}$ ions should be sufficiently large as
well the nominal charge difference. With the increase of La concentration, the extra doping electrons 
populate the Mo t$_{2g}$ down spin sub-band crossing the Fermi energy. As a result, the Mo$^{+5}$ 
nominal valence in SFMO becomes Mo$^{+5-m}$ in the doped compounds, $m$ being the number of doped 
electron with a maximum value of 2. This decreases the charge
difference between Fe$^{+3}$ and Mo. This is expected to be detrimental to the ordering, though the ionic
radii difference between Fe$^{3+}$ ( 0.645 \AA) and Mo$^{3+}$ (0.69 \AA) is larger than that
between Fe$^{3+}$ and Mo$^{5+}$ (0.65 \AA). The study on SFMO in this context \cite{sfmo-domain}, find that even for
a disordered sample, as given by X-ray study, the local ordering is maintained with a domain structure.
Annealing conditions can give rise to domain structures with varying sizes of the domain. This gives us the hope
in the observation of the antiferromagnetic phase in the La rich SFMO samples. Attempts are already being made to 
prepare these overdoped samples locally, and preliminary
results suggest existence of magnetic phases different from ferromagnetic phase~\cite{Sugatapriv}.
On theoretical front, effect of disorder has been studied by Alonso et. al. \cite{Guinea1}. 
Within a variational mean field framework, they found that the filling/doping at which the Tc goes to zero 
increases upon increasing antisite disorder. This means that antisite disorder stabilizes the ferromagnetic phase.
In other words, the antisite disorder is expected to reduce 
the stability of the antiferromagnetic phase, which is also seen in our preliminary calculations. We wish to carry out systematic 
study of the antisite disorder in future, keeping in mind possible formation of domain structures.

Finally, within the kinetic energy driven mechanism, the ferromagnetism gets destabilized and the
antiferromagnetism wins when the carrier concentration reaches such a value that the B$^{\prime}$ d states gets 
filled up in one spin channel and tries to populate the other spin channel. Such situation is encountered
also in case of another double perovskite, Sr$_{2}$CrOsO$_6$. Os being in nominal 5+ state, is in
d$^{3}$ configuration with completely full Os t$_{2g}$ states in the down spin channel and lies within
exchange split energy levels of Cr-t$_{2g}$\cite{MO}, a case very similar to La$_{2}$FeMoO$_6$.  Sr$_{2}$CrOsO$_6$,
in contrast to above expectation, however stabilizes in ferromagnetic configuration of Cr spins. In this context, we 
found that the delicate balance between FM and AFM, is governed by the extent of hybridization between the
localized B site and delocalized B$^{\prime}$ site. For Sr$_{2}$CrOsO$_6$, due to the movement of the Os t$_{2g}$
within the exchange split energy window of Cr-t$_{2g}$ in comparison to that of Mo t$_{2g}$
within the exchange split energy window of Fe-d, the hybridization effect is weakened and also the finite
spin-orbit (SO) coupling at Os site mixes the up and down spin channels, causing possibly 
the energy gain due to antiferromagnetism to be reduced\cite{MO}.

We believe our study will stimulate further experimental activities to explore the possibilities of double perovskites
exhibiting kinetic energy driven antiferromagnetism.

\section{Acknowledgment}
The authors gratefully acknowledge discussions with D. D. Sarma, P. Majumdar, M. Azuma and J. Gopalakrishnan. 
TSD acknowledges the support of Advanced Materials Research Unit and Swarnajaynti grant.


\newpage

\begin{thebibliography}{99}
\bibitem{SFMO}K.-I. Kobayashi, T. Kimura, H. Sawada, K. Terakura, and Y. Tokura, Nature (London) {\bf 395}, 677 (1998),                                                                                                                                                                                                                                                                                                            
\bibitem{BNMO}Masaki Azuma, Kazuhide Takata, Takashi Saito, Shintaro Ishiwata, Yuichi Shimakawa, and Mikio Takano, J. Am. Chem. Soc {\bf 127}, 8889(2005).
\bibitem{LNMO}  N. S. Rogado, J. Li, A. W. Sleight, and M. A. Subramanian, Adv. Mater. {\bf 17}, 2225 (2005). 
\bibitem{LNMO1}Hena Das, Umesh V. Waghmare, T. Saha-Dasgupta, and D. D. Sarma, Phys. Rev. Lett {\bf 100} 186402 (2008).
\bibitem{SCOO}Hena Das, Molly De Raychaudhury, and T. Saha-Dasgupta, Appl. Phys. Lett. {\bf 92}, 201912 (2008).
\bibitem{DDDinesh} J.B. Phillip, P. Majewski, L. Alff, A. Erb, R. Gross, T. Graf, M.S.Brandt, J. Simon, T. Walther, W. Mader, D. Topwal, 
and D.D. Sarma, Phys. Rev.B, {\bf 68}, 144431 (2003). 
\bibitem{DDreview} D.D. Sarma, Current Opinion in Solid State and Materials Science, {\bf 5},
261 (2001).
\bibitem{Mag-Res1}B. Garcia Landa et al, Solid State Comm., {\bf 110}, 435 (1999).
\bibitem{Mag-Res2} B. Martinez, J. Navarro, L. Balcells and J. Fontcuberta, J. Phys.: Condens. Matter {\bf 12}, 10515 (2000).
\bibitem{Mag-Res3}D. D. Sarma, S. Ray, K. Tanaka, M. Kobayashi, A. Fujimori, P. Sanyal, H.R. Krishnamurthy and C. Dasgupta,
Phys. Rev. Lett., {\bf 98}, 157205 (2007).
\bibitem{Mag-Res4}C. L. Yuan et al, Appl. Phys. Lett. {\bf 75}, 3853 (1999)
\bibitem{Mag-Res5}D.D. Sarma, E.V. Sampathkumaran, R. Sugata, R. Nagarajan, S. Majumdar, A. Kumar, G. Nalini and T.N. Gururow. Solid State Commun. {\bf 114}, 465 (2000). 
\bibitem{kaputza} Cz. Kapusta, P. C. Riedi, D. Zajac, M. Sikora, J. M. De Teresa, L. Morellon, M. R. Ibarra, J. Magn. Magn. Materials, {\bf 242-245}, 701 (2002).
\bibitem{kuepper} K Kuepper, M Kadiroglu, A V Postnikov, K C Prince,
M Matteucci,V R Galakhov, H Hesse, G Borstel and M Neumann, J. Phys.: Condens. Matter {\bf 17} 4309 (2005).
\bibitem{tomika} Y. Tomioka, T. Okuda, Y. Okimoto, R. Kumai, K.-I. Kobayashi, Y. Tokura, Phys. Rev. B {\bf 61}, 422 (2000).
\bibitem{topwal} Dinesh Topwal, D. D. Sarma, H. Kato, Y. Tokura and M. Avignon, Phys. Rev. B {\bf 73}, 094419 (2006).
\bibitem{TSD}D.D. Sarma, P. Mahadevan, T. Saha Dasgupta, S. Ray, A. Kumar, Phys. Rev. Lett., {\bf 85}, 2549 (2000).
\bibitem{Szotek}Z. Szotek, W. M. Temmerman, A. Svane, L. Petit, and H. Winter, Phys. Rev. B {\bf 68}, 104411 (2003).
\bibitem{Delin}V. Kanchana, G. Vaitheeswaran, M. Alouani, and A. Delin, Phys. Rev. B {\bf 75}, 220404(R) (2007).
\bibitem{Solovyev} I. V. Solovyev, Phys. Rev. B {\bf 65}, 144446 (2002).
\bibitem{Millis}A. Chattopadhyay and A. J. Millis, Phys. Rev. B {\bf 64}, 024424 (2001).
\bibitem{Avignon}O. Navarro, E. Carvajal, B. Aguilar, M. Avignon, Physica B {\bf 384}, 110 (2006).
\bibitem{GK}Junjiro Kanamori and Kiyoyuki Terakura, Journal of the Physical Society of Japan {\bf 70}, 1433 (2001)
\bibitem{Navarro}E. Carvajal, O. Navarro, R. Allub, M. Avignon and B. Alascio, Eur. Phys. J. B, {\bf 48}, 179-187 (2005).
\bibitem{Guinea2}L. Brey, M. J. Calder$\acute{o}$n, S. Das Sarma and F. Guinea, Phys. Rev. B {\bf 74}, 094429 (2006).
\bibitem{Guinea1} J.L.Alonso,L.A. Fernandez, F. Guinea, F. Lesmes, and V. Martin-Mayor, Phys. Rev. B, {\bf 67}, 214423 (2003).
\bibitem{Prabs}Prabuddha Sanyal and Pinaki Majumdar, Phys. Rev. B, {\bf 80}, 054411 (2009). 
\bibitem{SFWO}Z. Fang, K. Terakura and J. Kanamori, Phys. Rev. B {\bf 63}, 180407 (2001).
\bibitem{Khomskii} J. van den Brink, G. Khaliullin, D. Khomskii, Phys. Rev. Lett., {\bf 83}, 5118 (1999). 
\bibitem{JAP}A. Kahoul, A. Azizi, S. Colis, D. Stoeffler, R. Moubah, G. Schmerber, C. Leuvrey, and A. Dinia, J. Appl. Phys. {\bf 104}, 123903 (2008).
\bibitem{ES}T. Saitoh, M. Nakatake, H. Nakajima, O. Morimoto, A. Kakizaki, Sh. Xu, Y. Moritomo, N. Hamada, Y. Aiura, Journal of Electron Spectroscopy and Related Phenomena {\bf 144-147}, 601 (2005).
\bibitem{VASP}G. Kresse and J. Hafner, Phys. Rev. B {\bf 47}, 558 (1993), G. Kresse and J. Furthmueller, Phys. Rev. B {\bf 54}, 11169 (1996).
\bibitem{NMTO}O. K. Andersen and T. Saha-Dasgupta, Phys. Rev. B {\bf 62}, R16219 (2000).
\bibitem{GGA}J. P. Perdew, J. A. Chevary, S. H. Vosko, K. A. Jackson, M. R. Pederson, D. J. Singh, and C. Fiolhais, Phys. Rev. B {\bf 46}, 6671 (1992); {\bf 48}, 4978(E) (1993).
\bibitem{LSDA+U}V. I. Anisimov, F. Aryasetiawan, and A. I. Liechtenstein, J. Phys.: Cond. Mat.{\bf 9}, 767 (1997).
\bibitem{PAW}P. E. Bl$\ddot{o}$chl, Phys. Rev. B {\bf 50}, 17953 (1994). G. Kresse and D. Joubert, Phys. Rev. B {\bf 59}, 1758 (1999).
\bibitem{lmto} O. K. Andersen, Phys. Rev. B, {\bf 12}(1975) 3060.
\bibitem{P21} J. Navarro, C. Frontera, Ll. Balcells, B. Mart$\acute{i}$nez, and J. Fontcuberta, Phys. Rev. B {\bf 64}, 092411 (2001).
\bibitem{p21n} Carlos Frontera, Diego Rubi, Jose Navarro, Jose Luis Garcia-Munoz, and Josep Fontcuberta, Phys. Rev. B {\bf 68}, 012412 (2003).
\bibitem{note} Carlos Frontera (private communication).
\bibitem{note_op} Optimization has been carried out both in terms of GGA and GGA+U. The results are found to differ only marginally. The values
quoted in Table I. were obtained with GGA.
\bibitem{photoemission} J. Navarro, J. Fontcuberta, M. Izquierdo, J. Avila, M.C. Asensio, Phys. Rev. B,{\bf 70},054423 (2004).
\bibitem{uvalue}Ze Zhang and Sashi Satpathy, Phys. Rev. B 44, 13319 (1991)
\bibitem{tl2mn2o7} T. Saha-Dasgupta, Molly De Raychaudhury, D. D. Sarma, Phys. Rev. Lett. {\bf 96}, 087205 (2006)
\bibitem{footnote} There are nine degrees of freedom per Fe-Mo pair, consisting of three at Fe site and six at Mo site.
\bibitem{sfmo-domain} C. Meneghini, Sugata Ray, F. Liscio, F. Bardelli, S. Mobilio, and D. D. Sarma,  Phys. Rev. Lett. 103, 046403 (2009).
\bibitem{Sugatapriv} S. Ray, {\em private communication}.  
\bibitem{MO}H. Das, P. Sanyal, T. Saha-Dasgupta, and D. D. Sarma (unpublished).


\end{thebibliography}
\end{document}